\documentclass[conference,unknownkeysallowed]{IEEEtran}

\IEEEoverridecommandlockouts
\def\BibTeX{{\rm B\kern-.05em{\sc i\kern-.025em b}\kern-.08emT\kern-.1667em\lower.7ex\hbox{E}\kern-.125emX}}

\usepackage[font={small, it}]{caption}    
\usepackage{todonotes}    
\usepackage{nicefrac}
\usepackage{siunitx}
\usepackage{array,framed}
\usepackage{booktabs}
\usepackage{fancyvrb}
\usepackage{graphicx}

\usepackage{
  color,
  float,
  epsfig,
  wrapfig,
  graphics,
  graphicx,
  subcaption
}
\usepackage{textcomp,amssymb}
\usepackage{setspace}
\usepackage{latexsym,fancyhdr,url}
\usepackage{enumerate}
\usepackage{algorithm2e}
\usepackage{algpseudocode}
\usepackage{xparse} 
\usepackage{xspace}
\usepackage{multirow}
\usepackage{csvsimple}
\usepackage{balance}

\newcommand{\susmit}{\color{black}}

\usepackage{
  tikz,
  pgfplots,
  pgfplotstable
}
\usepackage{hyperref}

\usetikzlibrary{
  shapes.geometric,
  arrows,
  external,
  pgfplots.groupplots,
  matrix
}

\pgfplotsset{compat=1.9}

\usepackage{mathtools,}

\DeclareMathAlphabet{\mathcal}{OMS}{cmsy}{m}{n}

\DeclareGraphicsExtensions{%
    .png,.PNG,%
    .pdf,.PDF,%
    .jpg,.mps,.jpeg,.jbig2,.jb2,.JPG,.JPEG,.JBIG2,.JB2}

\usepackage{xparse}
\newcommand{\bnm}{\begin{newmath}}
\newcommand{\enm}{\end{newmath}}

\newcommand{\bea}{\begin{eqnarray*}}%
\newcommand{\eea}{\end{eqnarray*}}%

\newcommand{\bne}{\begin{newequation}}
\newcommand{\ene}{\end{newequation}}

\newcommand{\bal}{\begin{newalign}}
\newcommand{\eal}{\end{newalign}}

\newenvironment{newalign}{\begin{align}%
\setlength{\abovedisplayskip}{4pt}%
\setlength{\belowdisplayskip}{4pt}%
\setlength{\abovedisplayshortskip}{6pt}%
\setlength{\belowdisplayshortskip}{6pt} }{\end{align}}

\newenvironment{newmath}{\begin{displaymath}%
\setlength{\abovedisplayskip}{4pt}%
\setlength{\belowdisplayskip}{4pt}%
\setlength{\abovedisplayshortskip}{6pt}%
\setlength{\belowdisplayshortskip}{6pt} }{\end{displaymath}}

\newenvironment{newequation}{\begin{equation}%
\setlength{\abovedisplayskip}{4pt}%
\setlength{\belowdisplayskip}{4pt}%
\setlength{\abovedisplayshortskip}{6pt}%
\setlength{\belowdisplayshortskip}{6pt} }{\end{equation}}

\newcounter{ctr}

\newcounter{mytable}
\def\mytable{\begin{centering}\refstepcounter{mytable}}
\def\endmytable{\end{centering}}

\newcounter{myfig}
\def\myfig{\begin{centering}\refstepcounter{myfig}}
\def\endmyfig{\end{centering}}

\newlength{\saveparindent}
\newlength{\saveparskip}
\newcommand{\E}{{\rm I\kern-.3em E}}

\renewcommand{\eqref}[1]{\mbox{Equation~(\ref{#1})}}

\def \part {part}

\renewcommand{\paragraph}[1]{\vspace*{6pt}\noindent\textbf{#1}\;}

\def \blackslug{\hbox{\hskip 1pt \vrule width 4pt height 8pt
    depth 1.5pt \hskip 1pt}}
\def \qed{\quad\blackslug\lower 8.5pt\null\par}

\newcounter{mynote}[section]

\newcommand\ignore[1]{}

\newcounter{rcnote}[section]

\newcounter{mrnote}[section]

\newcounter{fknote}[section]

\newcounter{anote}[section]

\DeclareMathSymbol{\mlq}{\mathord}{operators}{``}
\DeclareMathSymbol{\mrq}{\mathord}{operators}{`'}

\newcommand{\rhf}[2]{R_{f, \gamma}}

\DeclareDocumentCommand{\edist}{o o}{
  \ensuremath{
    \IfNoValueTF{#1}{{d}}{{\sf d}(#1,#2)}
  }
}

\newcommand{\olrk}[1]{\ifx\nursymbol#1\else\!\!\mskip4.5mu plus 0.5mu\left(\mskip0.5mu plus0.5mu #1\mskip1.5mu plus0.5mu \right)\fi}

\NewDocumentCommand{\indseq}{ O{1} O{r} }{{#1}\ldots {#2}}

\usepackage[absolute]{textpos}

\begin{document}

\begin{textblock*}{\textwidth}(3cm,1cm)
\noindent\colorbox{red}{This is the accepted version of the paper. The final version of this paper will appear at IEEE CloudNet 2020.}
\end{textblock*}

\fancyhead{}
\def\thetitle{ Named Data Networking for Content Delivery Network Workflows}
\title{\thetitle}

\author{
\IEEEauthorblockN{Rama Krishna Thelagathoti\IEEEauthorrefmark{1},
Spyridon Mastorakis\IEEEauthorrefmark{1},
Anant Shah\IEEEauthorrefmark{2},
Harkeerat Bedi\IEEEauthorrefmark{2} and
Susmit Shannigrahi\IEEEauthorrefmark{3}}
\IEEEauthorblockA{\IEEEauthorrefmark{1}Computer Science Department, University of Nebraska, Omaha, USA\\
Email: rthelagathoti@unomaha.edu, smastorakis@unomaha.edu}
\IEEEauthorblockA{\IEEEauthorrefmark{2}Verizon Digital Media, USA\\
Email: anant.shah@verizondigitalmedia.com,  harkeerat.bedi@verizondigitalmedia.com}
\IEEEauthorblockA{\IEEEauthorrefmark{3}Computer Science Department, Tennessee Tech University, Cookeville, TN\\
Email: sshannigrahi@tntech.edu}}



\maketitle
\IEEEpubidadjcol
\begin{abstract}

In this work we investigate Named Data Networking's (NDN's) architectural properties and features, such as content caching and intelligent packet forwarding, in the context of a Content Delivery Network (CDN) workflows. More specifically, we evaluate NDN's properties for PoP (Point of Presence) to PoP and PoP to device connectivity. We use the Apache Traffic Server (ATS) platform to create an HTTP, CDN-like caching hierarchy in order to compare NDN with HTTP-based content delivery. Overall, our work demonstrates that properties inherent to NDN can benefit content providers and users alike. Our experimental results demonstrate that HTTP is faster under stable conditions due to a mature software stack. 
However, NDN performs better in the presence of packet loss, even for a loss rate as low as 0.1\%, due to packet-level caching in the network and fast retransmissions from close upstreams and fast retransmissions from close upstreams. We further show that the Time To First Byte (TTFB) in NDN is consistently lower than HTTP ($\sim 100ms$ in HTTP vs $\sim50ms$ in NDN), a vital requirement for CDNs, in addition to supporting transparent failover to another upstream when a failure occurs in the network. 
Moreover, we examine implementation agnostic (implementation choices can be Software Defined Networking, Information Centric Networking, or something else) network properties that can benefit CDN workflows.

\end{abstract}


\section{Introduction} \label{sec:intro}

The primary goal of the current Internet is content distribution. Content Delivery Networks (CDNs) serve content to a large number of users, predominately over HTTP(s). HTTP over TCP/IP has worked well, though not without complex configuration and extreme engineering solutions.

Named Data Networking (NDN)~\cite{zhang2014named} is a future Internet architecture that focuses on content distribution and utilizes content names for all operations. The synergy between CDNs operations and NDN architecture is that both aim to optimize content delivery. Further, CDNs utilize named content addressed by HTTP URLs. Similarly, NDN utilizes named content for content both content delivery and in-network operations such as routing and forwarding. By aligning the underlying network to CDNs' requirements, we can expect several benefits - less complicated infrastructure and software stack, optimized data delivery, and automatic path optimization at the network layer.  {\susmit Initially, NDN can be deployed as an "overlay" on existing CDN infrastructure between Points-of-Presence (PoPs) that will not require any infrastructure change. After the benefits are well understood and evaluated at scale, native NDN deployment can be considered.}

In this work, we look at content distribution over HTTP using Apache Traffic Server (ATS), an open-source, high performance caching software that can act both as reverse and forward proxy. For looking at content distribution over NDN, we use the standard NDN software such as ndn-tools (producer/consumer tools) and NDN forwarding daemon (NFD)~\cite{afanasyev2014nfd}.  Using Ubuntu virtual machines (VMs) on Google Cloud Platform (GCP), we create a simple topology of nodes. 
Our goal is to evaluate {\susmit NDN's performance based on salient features of a CDN topology.}

Our contribution in this paper is twofold. First, we compare NDN with HTTP on an testbed of virtual machines. These comparisons demonstrate in which scenarios NDN performs better than HTTP. {\susmit The second is to identify and quantify the benefit that the NDN architecture can bring to a contemporary CDN architecture. We show that properties such as seamless failover, the ability to choose the ``best upstream", and partial data retrieval can benefit CDNs.}


\textbf{Motivation} The current TCP/IP paradigm is end-to-end. On the contrary, the CDNs aim to deliver content in a way that is transparent to the user. Creating a location transparent overlay on top of an end-to-end protocol stack requires hiding several end-to-end artifacts from the users. For example, when using TCP, end to end connections need to be established, even when the transfer size is small. Upstream hosts (caches and origins) need to be enumerated and monitored continuously, and the requests need to be routed to them. When an upstream host fails, a TCP session needs to restart, potentially throwing away already retrieved content. On the contrary, NDN automatically provides a hop-by-hop packet forwarding as well as location independent data retrieval. Since NDN's network model perfectly aligns with CDN's application model, our motivation is to explore these synergies. 

\textbf{Paper organization:} The rest of the paper is organized as follows - Section \ref{sec:relwork} presents a brief introduction to a contemporary CDN architecture and NDN. It also discusses previous work that attempted to either apply NDN to CDN or compare NDN to HTTP. Section \ref{sec:experiments} presents an overview of our experimental topology and setup. Section \ref{sec:results} evaluates NDN's architectural properties and points out the benefits of NDN. Section \ref{sec:conclusion} discusses our future directions and concludes.

\section{Background and Related Work}\label{sec:relwork}

\subsection {Named Data Networking}

Named Data Networking (NDN)~\cite{zhang2014named} is an instance of the Information Centric Networking (ICN) paradigm~\cite{xylomenos2013survey}. NDN utilizes application-defined, semantically meaningful, and hierarchical names for data publication. Once data is named and published, a user may send an Interest packet (query) into the network specifying the name of the requested content. The intermediate routers maintain a name-based forwarding table that performs longest prefix match on the Interest name and forwards it over the appropriate interface(s). Once a producer receives the Interest, it returns a digitally signed Data packet. Intermediate routers cache these Data packets for future requests. For brevity, we do not discuss the NDN architecture further but refer the reader to the relevant publications \cite{zhang2014named}\cite{afanasyev2014nfd}.


NDN utilizes a hop-by-hop communication primitive and not an end-to-end principle like TCP/IP. NDN also has one to one flow balance, that is, for one Interest packet, only one data packet is returned. A one MB file on the disk is broken into several smaller Data packets (typically 8800 bytes or less). The client will send out an Interest for each of these Data packets. This hop by hop principle and the one-to-one flow balance means that in-network routers can perform several operations that were not easy to perform with the traditional TCP/IP architecture. Functions such as transparent failover to another source, parallel data retrieval from multiple upstreams, and utilizing the ``best" path can greatly benefit CDN operations.

Further, since there is no end-to-end connection setup, data retrieval starts faster (time to the first byte) and helps with faster content delivery. Since the content is broken into smaller chunks, NDN supports partial data retrieval through caching; only the part that is not present in a cache can be retrieved from the origin server, reducing the total transfer volume.

\subsection{Previous work}

Yuan et al. presented an early study between HTTP and CCNx (an incarnation of the NDN concepts)~\cite{yuan2013experimental}. The work compares HTTP and CCNx and shows the in networks with  with very high drop rates (10\% or more) NDN outperforms HTTP. However, the results do not apply to CDNs since such high loss rates in a CDN infrastructure are unacceptable.

Jiang et al. present nCDN that embeds NDN into the existing CDN framework by utilizing a hybrid model that uses NDN over UDP/TCP for content delivery~\cite{jiang2014ncdn}. 
The work requires a proxy that translates requests from the browser into NDN Interests before sending them into an NDN network. nCDN aims to improve the consumer experience while we focus on improving the backend infrastructure (e.g., PoP to PoP and POP to the origin server).

Lertsinsrubtavee et al. present a study of CDN vs. NDN in a wireless mesh network~\cite{lertsinsrubtavee2014comparing}. The study is done for wireless networks with completely different characteristics (e.g., loss and delay). Ma et al. present a comparison between CDN and NDN~\cite{ma2014tentative}. They show that NDN greatly simplifies the CDN design, provides better bandwidth utilization, and makes the CDN more scalable. 
The work focuses solely on throughput and misses the more useful parameters for an actual CDN provider, such as time to first byte (TTFB) and partial content delivery from multiple caches.

Mastorakis et al. performed comparative studies between NDN-based and IP-based solutions in the context of edge computing~\cite{mastorakis2020icedge} as well as for data sharing in infrastructure-based~\cite{mastorakis2017ntorrent} and infrastructure-less~\cite{mastorakis2020dapes} environments. Finally, Shannigrahi et al. present several studies that apply NDN for scientific data delivery in a fashion similar to CDN~\cite{shannigrahi2017request, fan2015managing}. However, these studies did not focus on CDN scenarios.

This study focuses on finding NDN properties that can be beneficial to a contemporary CDN backend infrastructure.
\section{Methodology}\label{sec:experiments}

\begin{figure}[!ht]
    \centering
    \includegraphics[width=0.8\columnwidth]{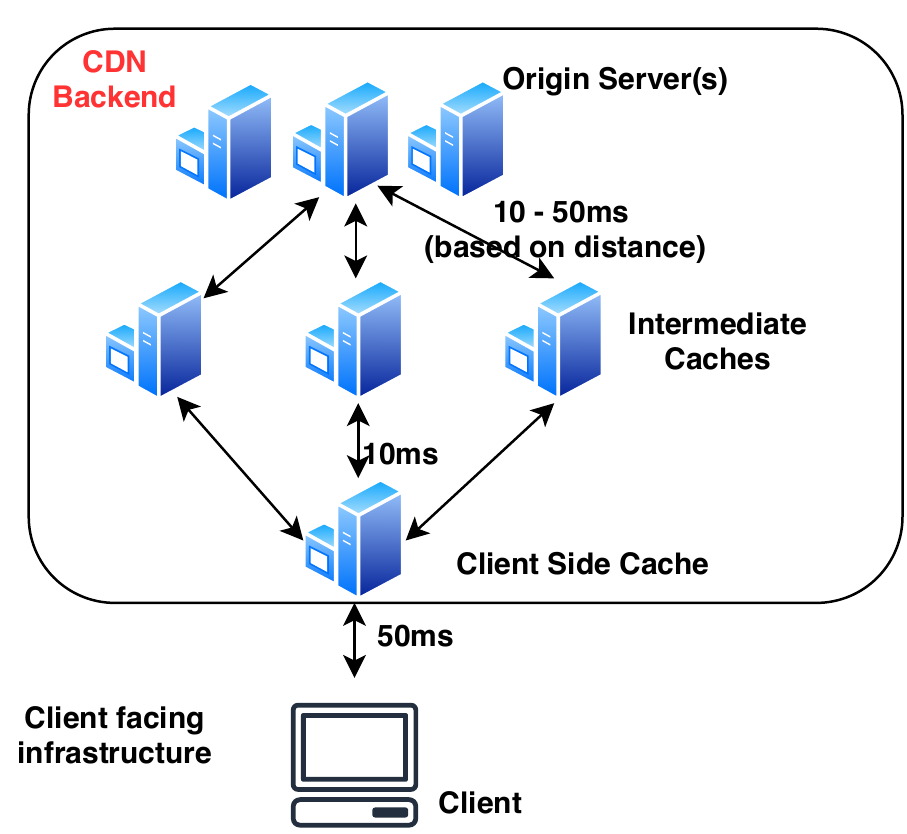}
    \caption{Experimental Topology. The infrastructure inside the box represents a CDN backend (PoP-to-PoP) - the focus of this work.}
    \label{fig:topology}
\end{figure}

Figure \ref{fig:topology} shows the topology we used for evaluating NDN for a CDN. While simplistic, this topology is sufficient for demonstrating NDN and HTTP's properties in a CDN context. We utilized Ubuntu virtual machines (VMs) deployed on the Google Cloud Platform (GCP) to perform these experiments. Each VMs had 4 CPUs, 4GB of RAM, and 10GB of disk space. The VMs were in the same region on Google cloud. Depending on the experiment, we utilize the linux utility ``tc" to set delay and latecy on the links (see Figure. \ref{fig:topology}).

{\susmit The client represents a device that requests content from the CDN. The client side cache can be deployed near the users to reduce latency and traffic volume. The intermediate caches provide hierarchical caching and origin shielding\cite{afergan2007method}. Any request that can not be served from the client side cache comes to these caches. These caches typically work as a reverse proxy, download the data from the origin, and caches it for future usage. The origin servers serve authoritative copies of the content. Depending on the use case, there might be one or more origin servers, located at different geographical locations. The client-side cache is 50 ms away from the client, the latency between other components are 10ms, and the latency between intermediate cache and origin servers are 10-50ms that enumerates distances to different origin servers.}



For the HTTP experiments, we utilize Apache Traffic Server (ATS)\cite{ATS}. CDN providers often use homegrown and performance-tuned HTTP servers for their services; ATS is the closet open-source alternative we used to create a caching hierarchy. In the topology (Figure. \ref{fig:topology}), we configured ATS on the client-side cache to act as a forward proxy - it simply forwards the requests of the clients to the server-side caches and caches the return data. The intermediate caches are configured as reverse proxies to the origin server. They receive the requests from the client-side cache, check if data is available in their local caches, and bring the data from the origin server before returning the data to the requesting entity (the client-side cache in this example). Multiple server-side caches are often utilized in a CDN to reduce the load on the caches and provide redundancy. 

For the HTTP experiments, we used Apache Web Server on the origin server as the content producer. On the intermediate caching nodes, we utilized HTTP caching with sizes set to 2-4GB. The content we requested were much smaller than the available caches sizes. We also manually mapped each ATS instance to their respective upstream nodes. In ATS, these mappings are static and must be configured before running experiments. Additionally, we had to set traffic as cachable using HTTP headers.


For failure and switching traffic between multiple upstream caches, we used ATS load balancer plugin. We used the round-robin policy that alternated traffic between two upstream servers. Note that failover and load balancing can also be done with DNS - we did not explore DNS failover in this work. Given the additional communication overhead needed for DNS, we anticipate the results to be worse. Finally, to retrieve content over HTTP, we used curl.

For the NDN experiments, we published data from the origin server using ndnputchunks, a standard ndn tool\cite{afanasyev2014nfd}. We added UDP overlays between the nodes for a namespace (e.g., /test) and requested content from the client. The intermediate caches were either disabled or set to a large number (2-4GB) depending on the experiment. We did not need any additional code for failover experiments since NDN supports it by default. For experiments with multiple upstreams, we added additional servers that had higher latencies. We ran each experiment at least ten times for statistical accuracy.

\section{Evaluation}\label{sec:results}

In this section, we discuss NDN and HTTP's performance in the context of a CDN. We compare the properties of HTTP over TCP (using ATS) with NDN. Note that HTTP over TCP, is highly developed and optimized over the last forty years, while NDN is still a research prototype. Given the difference in the quality of code and optimization, HTTP performs better in raw throughput numbers. However, our goal in this work is not to compare throughput (or goodput) but to point out NDN's unique properties that can benefit CDN operations. 

\subsection{Baseline Goodput}

\begin{figure}
    \centering
    \includegraphics[width=0.45\textwidth]{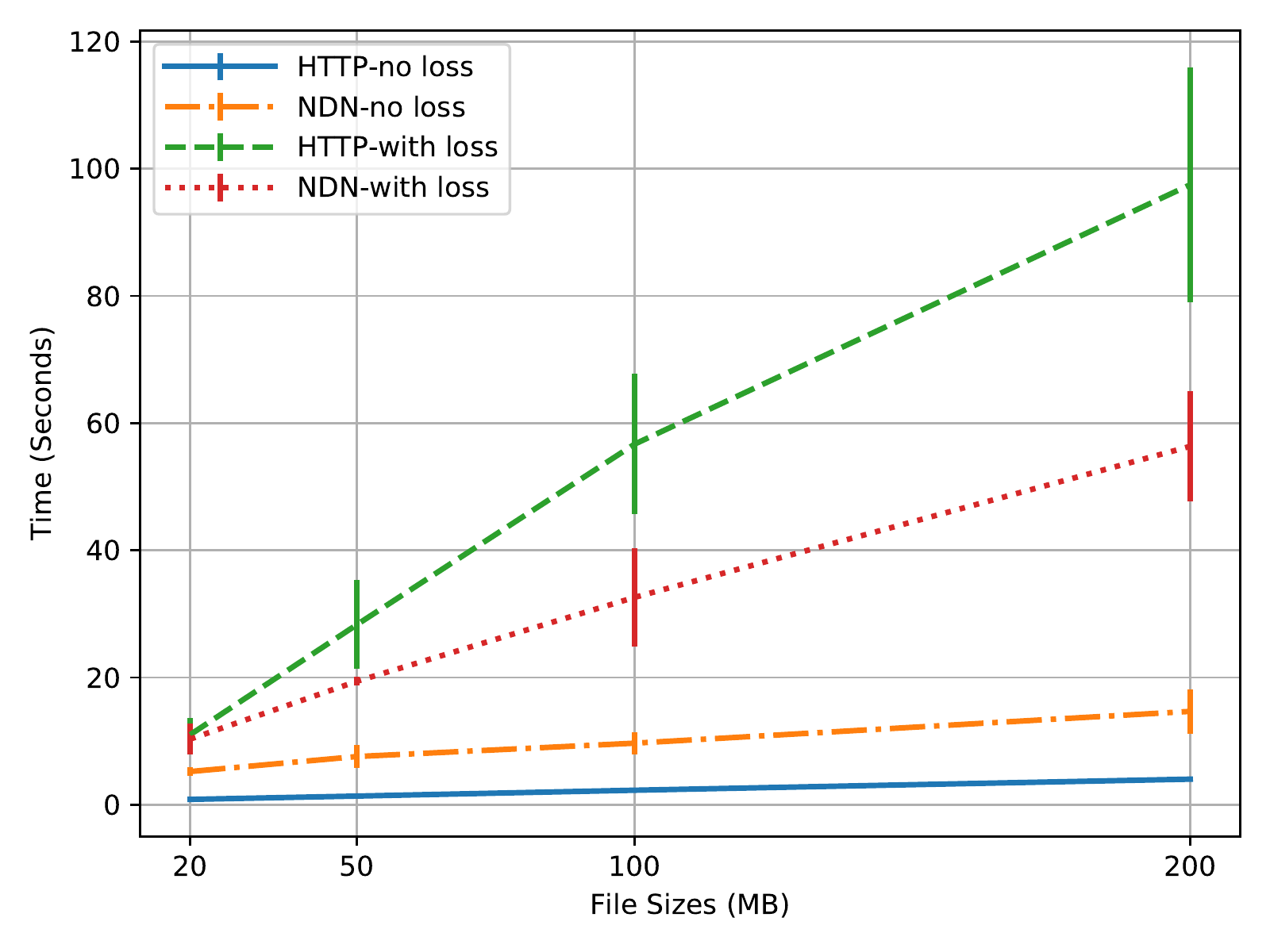}
    \caption{HTTP and NDN file retrieval performance. HTTP/ TCP is faster without any loss. In the presence of loss NDN performs better. The loss in this experiment was set to 0.08\% on the access link and 0.01\% on the other two links for a total of 0.1\% loss.}
    \label{fig:http_ndn}
\end{figure}

In this section, we compare NDN with ATS for goodput.  Figure \ref{fig:http_ndn} shows this difference - HTTP is consistently an order magnitude faster than NDN. However, the difference in performance is not exorbitant. For example, retrieving a 20 MB file from the origin server takes 0.1 seconds over HTTP, while the same file over NDN  takes 0.4 seconds. The numbers are similar when caching is enabled. Both NDN and HTTP times are proportionately lower with caching enabled. However, more importantly, the increase in delay follows the same pattern for both NDN and HTTP as the file size grows. The gap in performance observed can be reduced by careful performance tuning of NDN.

HTTP over TCP does not perform well in the presence of loss. Surprisingly, the percentage of loss does not have to be significant for NDN to outperform HTTP. In this experiment, we set the loss on the access link (between the client and the client-side cache) to 0.08\%. The other upstream links (between client-side cache and intermediate caches, and between the intermediate caches and the origin servers) a loss rate of 0.01\%. We performed these experiments with caching enabled.

When loss is present, NDN performs better as  retransmitted data comes from the client-side cache rather than from the origin. This allows the clients to get the data faster without going into a congestion avoidance mode. With HTTP, the connection is end-to-end, and packet loss means the TCP goes into congestion avoidance mode, slowing down the transfer. 




\subsection{Time To First Byte (TTFB)}

In this experiment, we measure the time between a request is sent and the client receives the first byte of data (TTFB) for HTTP and NDN.  TTFB is important for a CDN since it often dictates when the subsequent objects can be fetched (for example, various objects embedded in a webpage). It also affects the responsiveness of an application that can, in turn, affect the customer experience.

Our results indicate that TTFB is consistently lower for NDN compared to HTTP, with and without caching. Specifically, HTTP TTFBs are about 100 ms and 230 ms with and without caching, while NDN TTFBs are about 50 ms and 135 ms with and without caching, respectively. The reason for that is that HTTP runs on top of TCP that performs a handshake to establish a connection before retrieving any data. On the other hand, NDN can directly fetch data without requiring a connection creation or teardown, resulting in lower TTFB values. A lower TTFB is useful for delay-sensitive applications, for example, streaming videos.

\subsection{Cache Utilization}
\begin{figure}[!h]
    \centering
    \includegraphics[width=0.45\textwidth]{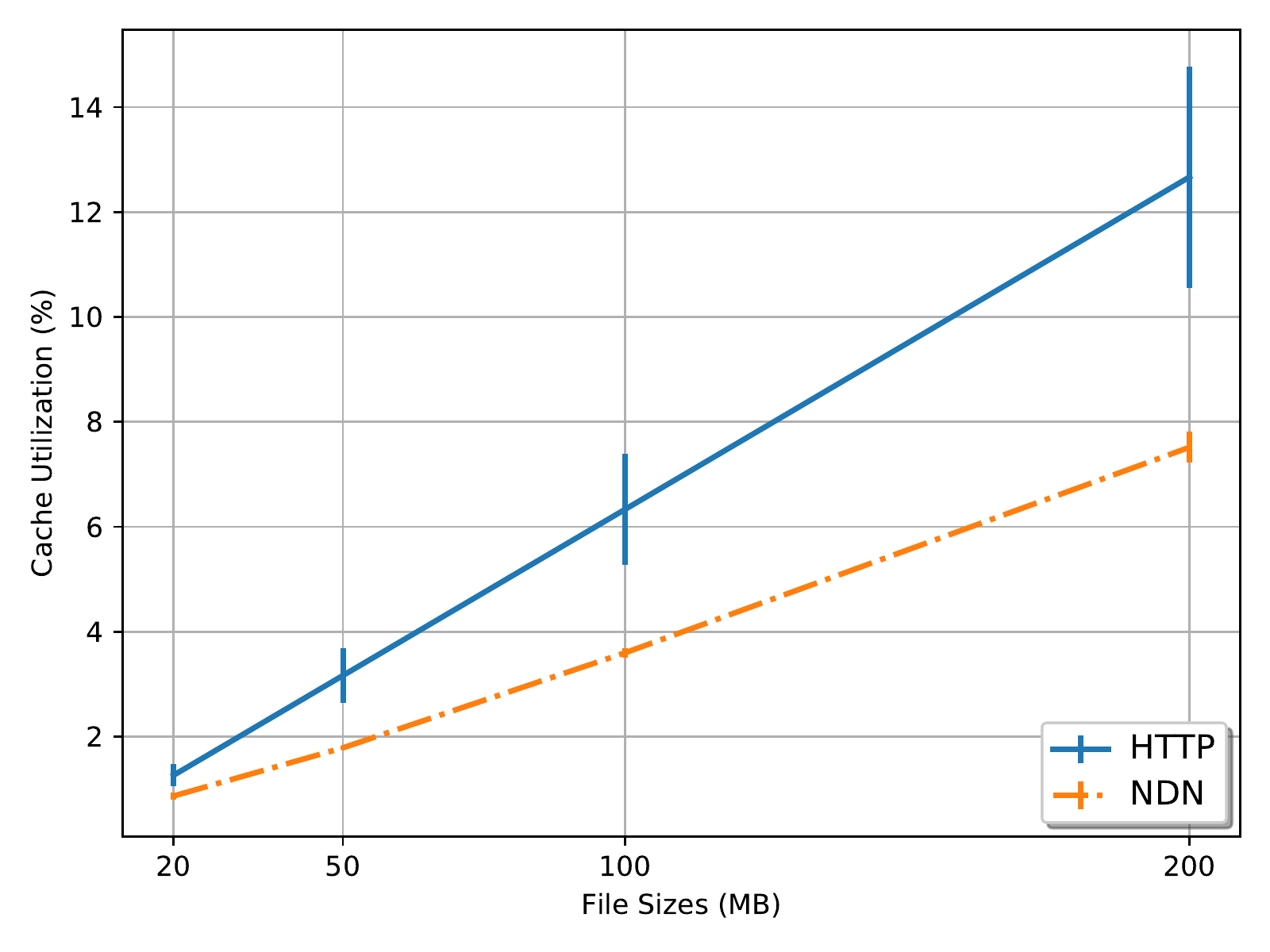}
    \caption{HTTP and NDN cache utilization}
    \label{fig:http_ndn_cache_utilization}
\end{figure}

CDNs are systems with cache hierarchies of large capacities. Maintaining a lower cache utilization per object across the system allows the CDNs to fit more content in the caches, speeds up delivery of a diverse set of content. However, as CDNs start to retrieve and cache content, cache utilization per content goes up. For example, if a piece of content is retrieved through one upstream cache and later the system uses another upstream, the content is now cached at two places. The effective cache utilization by this content is then   $\frac{2 \times \textrm{content size}}{\textrm{total\ cache\ size}}$.

With NDN, the caching granularity is packets. If half of the content is fetched from upstream cache one and the rest from upstream cache two, the utilization remains at  $\frac{\textrm{content size}}{\textrm{total\ cache\ size}}$.

Figure. \ref{fig:http_ndn_cache_utilization} shows this experiment. We start with a large cache for both NDN and HTTP that is much larger than the content we are retrieving. We enable caching for the intermediate caches but not on the origin and client-side cache. We start by retrieving a certain sized content (e.g. 100 MB) though intermediate cache 1, and in the middle of the retrieval, switch to the second source. If the switch happens after 10MB data is retrieved, after the second pull, NDN consumes  10MB at Intermediate cache one and 100MB at intermediate cache two (10 MB from the first pull + 100 MB from the second pull) in total. However, HTTP consumes 200MB of cache space across Intermediate cache one and Intermediate cache two since it must bring in the complete file.


\subsection{Partial Data Retrieval from Cache}
\begin{figure}[!h]
    \centering
    \includegraphics[width=0.45\textwidth]{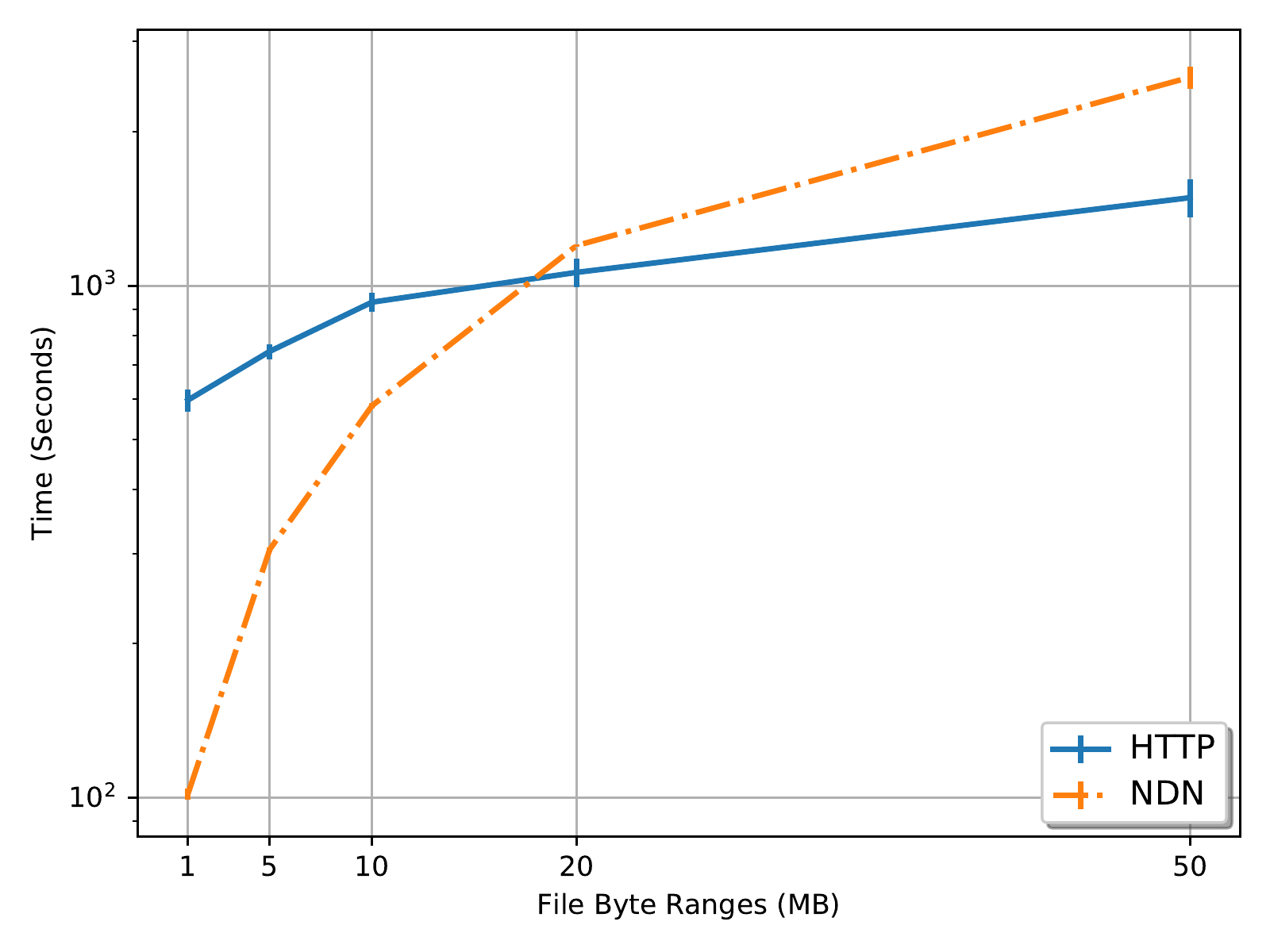}
    \caption{HTTP and NDN Partial Data Retrieval. The x-axis shows the byte ranges requested over HTTP and NDN. ATS either needs to fetch the full file or send the request to the origin server before serving a byte range request. NDN can reuse content chunks that are already present in the cache.}
    \label{fig:http_ndn_byte_range}
    
\end{figure}

In this experiment, we retrieve partial content from intermediate caches. With ATS, there are two ways to accomplish this. By default, the requests bypass the ATS caches and are directly forwarded to the origin server. Configuring ATS this way can negate origin shielding. A user can send several requests that request a sizable portion of a large file. The other way ATS supports byte range requests is by bringing in the whole file in the cache from the origin server and then serve the byte range requests from there. However, this approach is also problematic since requesting a small byte range from a large file will bring in the whole file into ATS's cache, reducing cache utilization.
On the contrary, NDN creates smaller chunks from a larger content object. For example, a large file (e.g., /data\_file) would be chunked into $n$ data packets, where the packets will have names like /data\_file/segment=1, /data\_file/segment=2,.../data\_file/segment=3. These segments are individually cached in the intermediate caches. An application asking for segments 1 to 10 can retrieve them from the cache without bringing the whole file in or going to the origin server. When partial data is cached, NDN can serve the subsequent requests from the cache - request for any content that is not in the cache to the origin server. As a result, NDN offers resiliency against both the scenarios in ATS where a user can potentially bypass the caches and reach the origin server or force ATS to bring in a large file.

Figure. \ref{fig:http_ndn_byte_range} shows the comparison of HTTP vs NDN Byte range requests. NDN is consistently faster than HTTP for a lower volume of data.  NDN outperforms ATS for partial data sizes up to  20MB. However, as the size of the byte range increases, NDN becomes slower compared to HTTP since it is not optimized for performance.

\subsection{Transparent Failover}

\begin{figure}[!h]
    \centering
    \includegraphics[width=0.45\textwidth]{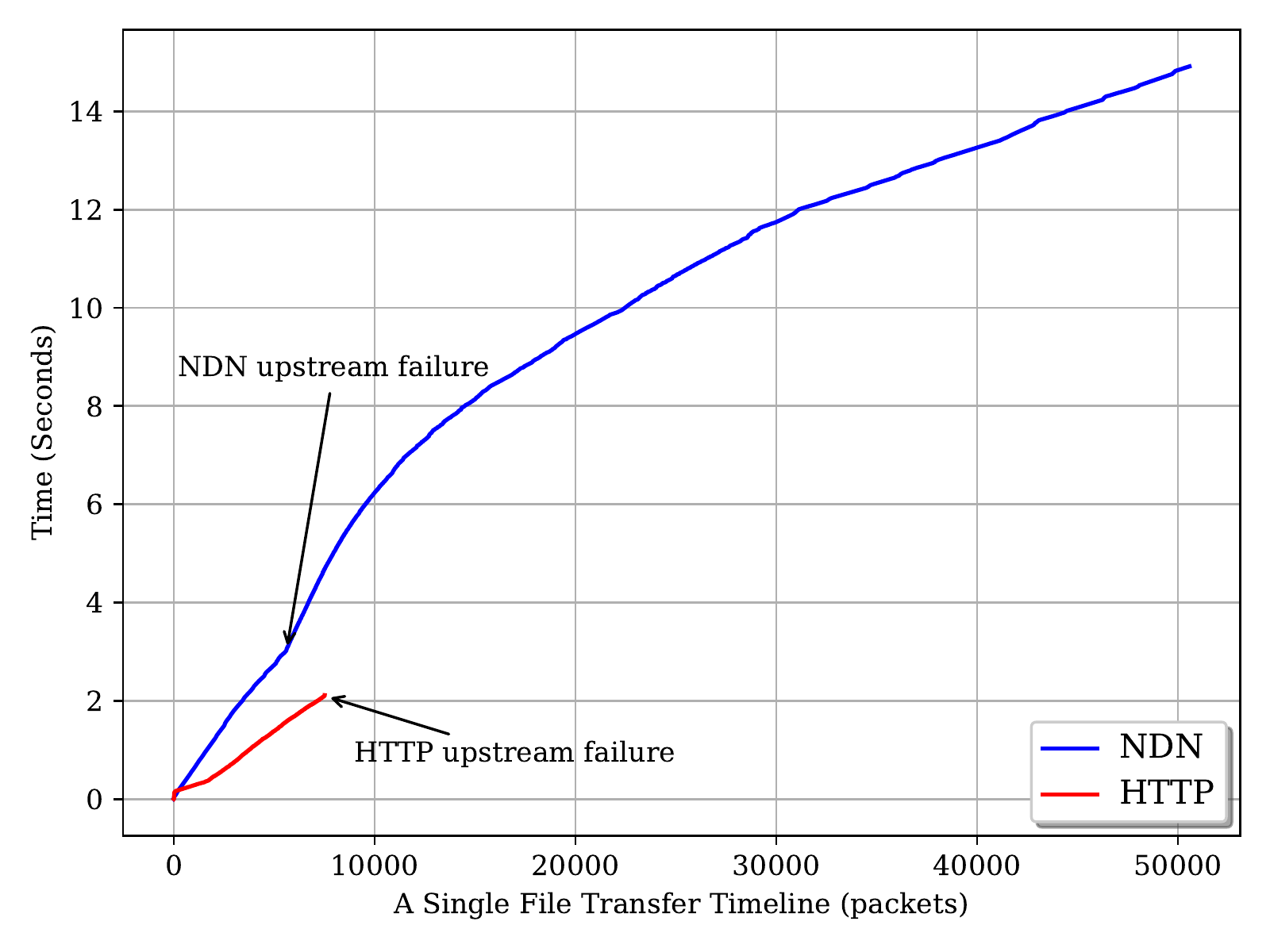}
    \caption{HTTP and NDN file retrieval perfomance with failure. NDN's hop by hop forwarding  transparently switches routes when upstreams fail. On the contrary, ATS connections are broken when upstream failure happens, throws away already retrieved content, and forces the client to restart the transfer. While currently HTTP is faster than NDN, this gap will close as NDN software stack matures.}
    \label{fig:http_ndn_failover}
\end{figure}




In a CDN, there are multiple intermediate caches (a hierarchy of caches). With HTTP and NDN, a user or a cache can be redirected to a better upstream based on network condition or the health of upstreams. However, in HTTP, the existing data flows stick to the same degraded path (TCP is end to end) even when multiple paths are available. When the upstream fails, the already retrieved data is thrown away, and the client has to restart the retrieval. On the contrary, NDN forwards packets hop by hop - it can switch upstreams in the middle of a transfer without restarting the retrieval.

In this experiment, we configured both ATS and NDN with multiple upstream servers. We pointed the ATS instance on the client-side-cache to two upstreams ATS instances that were acting as reverse proxies for the origin server. Similar to ATS, we pointed the NDN instance on the client-side-cache to two upstream NDN caches (on intermediate nodes 1 and 2). First, we started the transfer from the client, and as data starts to come in, find and kill the intermediate node that was serving the data. For example, the first client (HTTP) request went through the following path  - Client $\xrightarrow[]{}$ Client-Side Cache $\xrightarrow[]{}$ Intermediate Cache 1 $\xrightarrow[]{}$ Origin Server. When data started coming back, we stopped the ATS instance on Intermediate cache 1. Note that while ATS was configured with multiple upstreams, TCP connections are end-to-end, requiring reconnect on failure. In this scenario, the content retrieval stopped after failure occurred. 

We followed the same steps with NDN. We started the data transfer and killed the Intermediate node serving the request (the upstream), the client-side-cache immediately switched over to another available route, and there was no interruption in the transfer. The delay between two subsequent packets during the failover was 5.72ms with a standard deviation of 3.15ms. Figure \ref{fig:http_ndn_failover} demonstrates this scenario. The HTTP  transfer and the NDN transfer failed around the same time (at second 3 during the transfer). However, the NDN transfer continued with the alternate upstream while the HTTP transfer failed. Note that the NDN upstream switching is configurable and can happen without failures. NDN can be made to switch paths based on observed delay, loss, and any other relevant parameters to the CDN as we will show in the next section.

\subsection{Automatically switch to the better path}

\begin{figure}[!h]
    \centering
    \includegraphics[width=0.45\textwidth]{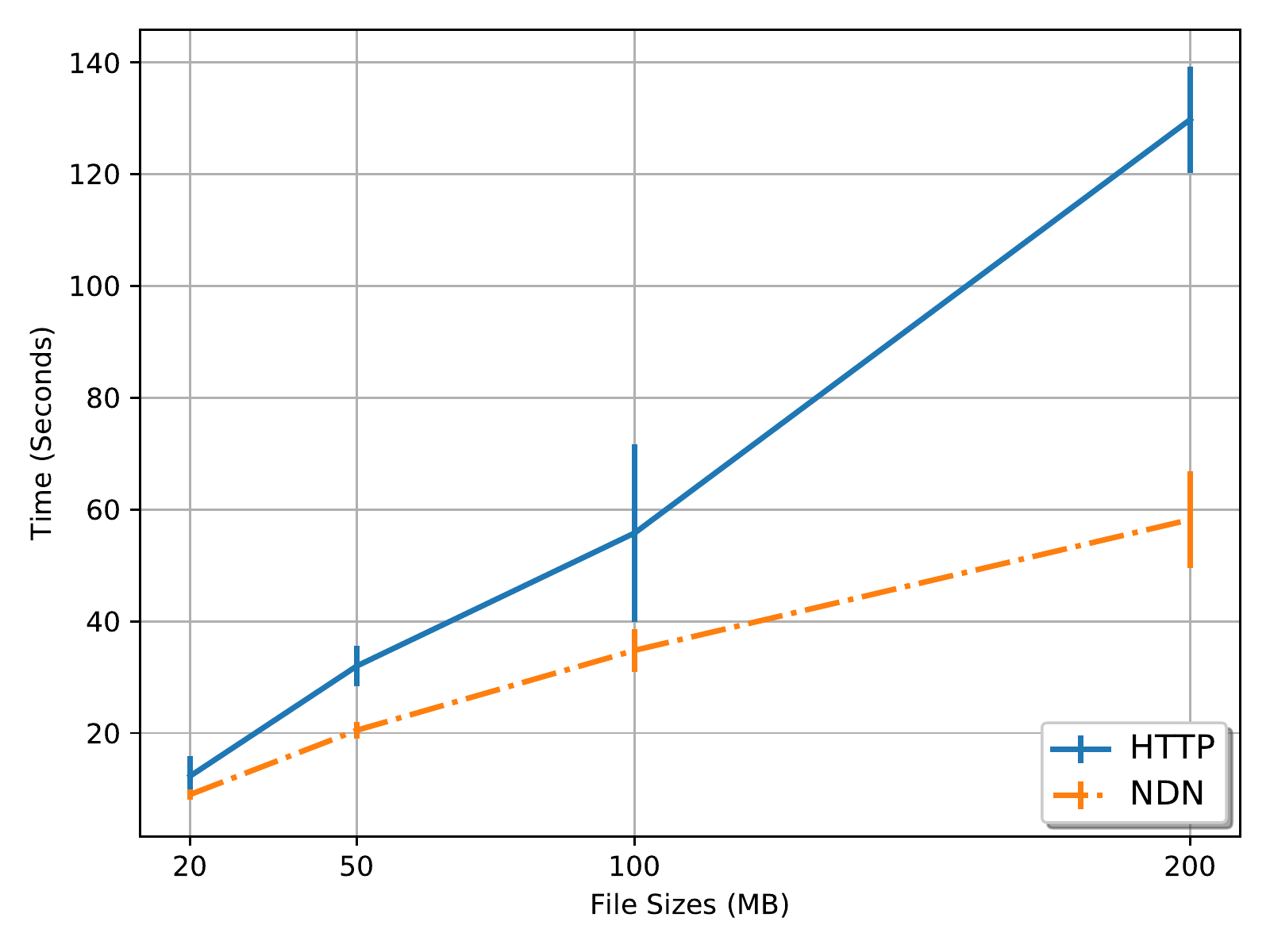}
    \caption{NDN and HTTP performance with upstream performance degradation with different file sizes. This experiment uses multiple upstreams. As the upstream detoriates, NDN moves to a better path while ATS sticks to the original path. The effect is more prominent as the file size increases.}
    \label{fig:route_switching}
\end{figure}

\begin{table*}[!ht]
\caption{Summary of the experiments and our observations}
    \label{tab:summary}
    \centering
    \begin{tabular}{|p{3cm}|p{15cm}|}
    \hline
        Experiment & Observations \\
    \hline
        A. Content retrieval & With lossy links, NDN retransmits faster from cache, resulting in better throughput. Without loss, HTTP has better throughput due to optimizations\\
    \hline
        B. TTFB & NDN follows a hop-by-hop model, TTFB is faster since no end-to-end connection set up is required \\
    \hline
        C. Cache Utilization & NDN provides packet level caching. Overall cache utilization is lower since the caching granularity is packets and not files\\
    \hline
        D. Partial retrieval & NDN provides packet level caching. Any data packet that is in the cache is reusable and the rest can be fetched from the origin. The ability to reuse packets from cache means NDN is faster when partial content is requested. \\
    \hline
        E. Failover & NDN can support multiple upstreams without special configurations. Combined with hop-by-hop forwarding, NDN can seamlessly switch over to another upstream when the current upstream fails. \\
    \hline
        F. Path switching & NDN strategies can monitor path parameters (defined by network operators) and automatically switch to better paths when current paths degrade.  \\
        
    \hline
    \end{tabular}
    
\end{table*}

In this experiment, we pick a ``best" link depending on the network condition. NDN's hop by hop forwarding allows network operators to monitor network performance and seamlessly move over to better data sources or links. Compare this to TCP flows, where performance degradation will require setting up a new TCP connection and restart the data retrieval. By not throwing away already retrieved content, NDN can improve completion time and provide a better experience for clients (they do not see the losses or quality degradation). In the context of a CDN, this is not only applicable to Client to PoP but also Pop to Pop communications.

In this experiment, we utilize two upstreams from the client-side cache.  For both NDN and HTTP, we configure upstreams routes with similar weights. The weight is a function of delay and loss. For simplicity, we choose the function to be (arbitrarily) $f(n) = \lceil 100* \textrm{delay\ in\ ms} + 100 * \textrm{percent\ packet\ loss} \rceil$. In the middle of data retrieval, we randomly change delay and loss. The loss is randomly chosen between 0.001\% and 1\%. The delay is somewhere between 50-200ms. With a 50 ms delay and 0.001\% loss, the weight of the path is 5000. When the delay increases to 100ms and the loss to 1\%, the weight becomes 5100. Note that the actual numbers are not important in this experiment as long as the weight of a route increases with the loss and delay, allowing us to switch over to a ``better route." In the context of a CDN, various other parameters 
can be added to the equation along with their respective weights.

With HTTP, an existing transfer sticks to the path even when the path deteriorates. ATS provides a mechanism to transparently switch over to another upstream by rewriting its configuration file. This mechanism also requires a seperate monitoring service that rewrites the ATS configuration file to direct traffic to a more preferred origin server. However, switching the upstream origin does not switch flows that are already transferring data. These flows will stick to the worse path until the transfers finish or restart the transfers.

With NDN, the data plane monitors the path quality and automatically switches over to an alternate path without involving any control plane mechanism. Additionally, by not throwing away already retrieved content, NDN can improve completion time and provide a better experience for clients (they don't see the losses or quality degradation).








\section{Conclusion and Future Work}
\label{sec:conclusion}
In this work, we demonstrate how NDN's architectural artifacts can help a CDN infrastructure. Table \ref{tab:summary} summarizes these observations. We utilize ATS to emulate a CDN-like caching hierarchy to compare HTTP based content delivery to NDN based content delivery. We show that NDN can deliver better performance in the presence of loss (even when the loss is small), reduce per-content cache utilization, and provide better time to the first byte, improving user experience. Further, NDN enables several mechanisms that are difficult to achieve using HTTP. Properties such as partial data retrieval, the ability to automatically switch to a better path, transparent failover, and the ability to retrieve content from multiple sources are important properties that can not only simplify a CDN infrastructure but also enhance user experience.

We acknowledge this work is preliminary. We plan to extend this study in several ways. First, we plan to utilize a representative CDN network topology with a larger set of links and nodes. We also plan to utilize real data access patterns observed in a CDN to investigate how NDN can benefit users and the CDN provider in terms of better resource utilization and efficient content delivery. Finally, we plan to create a pilot deployment inside a large scale CDN to understand better the trade-offs of applying a new Internet architecture to a very large-scale, complex, and distributed system. Finally, we plan to use a more performant version of the NDN forwarder, ndn-dpdk\cite{khoussi2019performance} that is under active development and can reach upto 100Gbps forwarding speed.



\bibliographystyle{IEEEtran}  
\bibliography{bib}


\end{document}